\documentclass[lettersize,journal]{IEEEtran}
\usepackage{verbatim}
\ifCLASSINFOpdf
\else
   \usepackage[dvips]{graphicx}
\fi
\usepackage{multirow}
\usepackage{booktabs}
\usepackage{caption}
\usepackage{graphicx}
\usepackage{epstopdf}
\usepackage{subfigure}
\usepackage{amsmath}
\usepackage{amssymb}
\usepackage{amsthm}
\usepackage{amsfonts}
\usepackage[linesnumbered,boxed,ruled,commentsnumbered]{algorithm2e}
\usepackage{algpseudocode}
\usepackage{array}
\usepackage{cite}
\usepackage{url}
\usepackage{color}
\usepackage{epstopdf}
\usepackage{overpic}
\usepackage{lipsum}
\usepackage{cuted}
\usepackage{stfloats}
\usepackage{amsxtra}
\usepackage{threeparttable}
\usepackage{mathrsfs}
\usepackage{ulem}
\newcommand{\bm}[1]{\mbox{\boldmath{$#1$}}}
\begin{document}
\title{Symmetric Saliency-based Adversarial Attack To Speaker Identification}

\author{Jiadi Yao, Chengdong Liang, Xing Chen, Xiao-Lei Zhang, Wei-Qiang Zhang, and Kunde Yang
\thanks{Corresponding author: Xiao-Lei Zhang}
\thanks{Jiadi Yao, Chengdong Liang, Xing Chen, Xiao-Lei Zhang, and Kunde Yang are with the School of Marine Science and Technology, Northwestern Polytechnical University, Xi'an 710072, China (e-mail: yaojiadi@mail.nwpu.edu.cn; liangchengdong@mail.nwpu.edu.cn;xing.chen@mail.nwpu.edu.cn; xiaolei.zhang@nwpu.edu.cn; ykdzym@nwpu.edu.cn).}
\thanks{Wei-Qiang Zhang is with the Department of Electronic Engineering, Tsinghua University, Beijing 100084, China (e-mail: wqzhang@tsinghua.edu.cn ).}
}

\markboth{Journal of \LaTeX\ Class Files,~Vol.~14, No.~8, August~2021}%
{Shell \MakeLowercase{\textit{et al.}}: A Sample Article Using IEEEtran.cls for IEEE Journals}


\maketitle

\begin{abstract}
Adversarial attack approaches to speaker identification either need high computational cost or are not very effective, to our knowledge. To address this issue, in this paper, we propose a novel generation-network-based approach, called symmetric saliency-based encoder-decoder (SSED), to generate adversarial voice examples to speaker identification. It contains two novel components. First, it uses a novel saliency map decoder to learn the importance of speech samples to the decision of a targeted speaker identification system, so as to make the attacker focus on generating artificial noise to the important samples. It also proposes an angular loss function to push the speaker embedding far away from the source speaker. Our experimental results demonstrate that the proposed SSED yields the state-of-the-art performance, i.e. over 97\% targeted attack success rate and a signal-to-noise level of over 39 dB on both the open-set and {close-set} speaker identification tasks, with a low computational cost.
\end{abstract}

\begin{IEEEkeywords}
Adversarial attack, speaker identification, saliency map decoder, angular loss.
\end{IEEEkeywords}

\section{Introduction}
The adversarial attack to speaker identification aims to make an identification system wrongly recognize the adversarial voice of a source speaker as a targeted imposter speaker, where the adversarial voice, a.k.a. \textit{adversarial example}, is produced by adding human-imperceptible noise to the speech of the source speaker. It shows great threat to modern speaker identification systems based on deep learning.
Existing adversarial noise generation methods can be categorized roughly into: {(i) gradient-based approaches \cite{kurakin2018adversarial, jati2021adversarial}, such as FGSM \cite{goodfellow2014explaining, li2020adversarial} and BIM \cite{kurakin2016adversarial}, (ii) optimization-based approaches, such as the C$\&$W attack \cite{carlini2017towards, joshi2021study}, Quasi-Newton \cite{goto2020quasi}, FoolHD \cite{shamsabadi2021foolhd} and AdvPulse \cite{li2020advpulse}, (iii) query-based approaches \cite{chen2021real}, and (iv) generation-network-based approaches, such as the Universal Adversarial Perturbations (UAPs) \cite{li2020universal} and FAPG \cite{xie2021enabling}. }

 The first three classes of the aforementioned methods are able to generate adversarial examples effectively with the expense of high computational cost, since that they need to search the optimal perturbation for each test utterance. To address the issue, the generation-network-based methods were proposed, which are able to generate an adversarial example of a test utterance in a single forward inference pass. However, to our knowledge, their performance might not be as high as the other classes in terms of targeted attack successful rate (TASR) and signal-to-noise ratio (SNR).

 To generate adversarial attacks to speaker identification efficiently that are able to achieve both high TASR and high SNR, in this paper, inspired by \cite{lu2021discriminator}, we propose a generation-network-based method, named Symmetric Saliency-based Encoder-Decoder (SSED), for the targeted attack to a speaker identification system. It has two novel components:
 \begin{itemize}
 \item A novel \textit{saliency map decoder} makes the attacker focus on the important samples of a test utterance that will strongly affect the decision of the targeted speaker identification system.
 \item A novel \textit{angular loss} makes the speaker embedding of an adversarial example far apart from the source speaker, which is a supplement to the mainstream of making the adversarial example close to the targeted speaker.
 \end{itemize}
Experimental results on both {close-set} speaker identification (CSI) and open-set speaker identification (OSI) \cite{chen2021real,zhang2022imperceptible} show that the proposed method achieves comparable performance to representative gradient- and optimization-based approaches with much higher efficiency than the latter. It also significantly outperforms a representative generation-network-based method with similar efficiency.

\begin{figure*}[t]
\begin{center}
\includegraphics[width=0.92\textwidth]{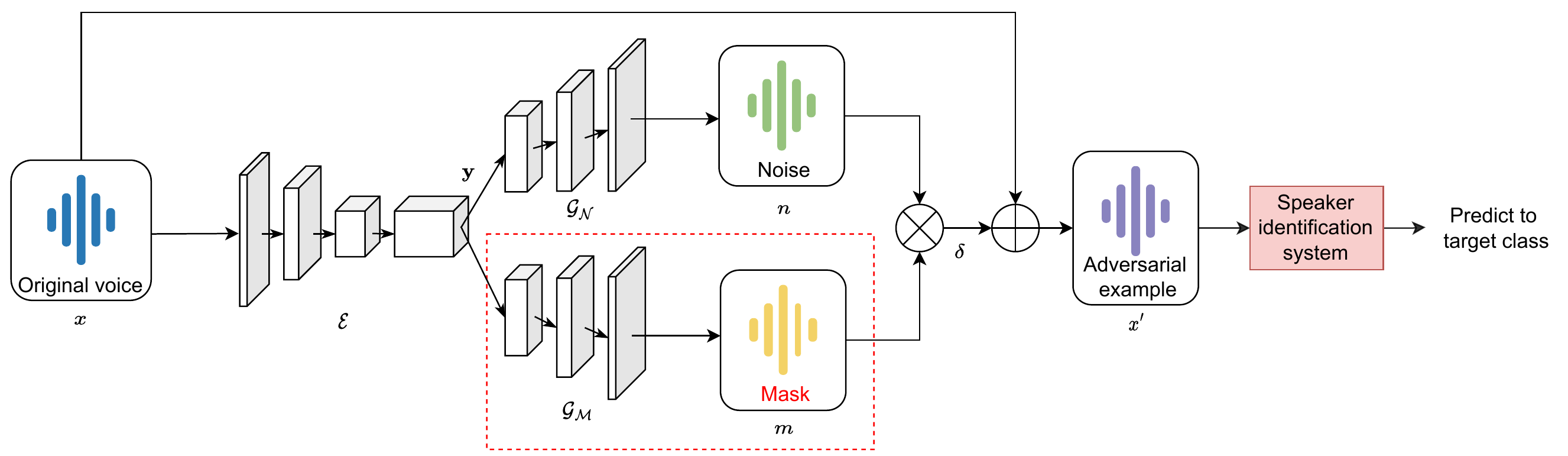}
\end{center}
\caption{Overview of the proposed symmetric saliency-based encoder-decoder. The saliency map decoder is in the red dot box.}
\label{fig:SSED}
\end{figure*}

\section{Preliminaries}
Speaker identification aims to detect the speaker identity of a test utterance from an enrollment database \cite{bai2021speaker}. If the speaker identity can never be out of the enrollment database, then it is a {close-set} identification (CSI) problem; otherwise, it is an open-set identification (OSI) problem. A speaker identification system contains an enrollment phase and a test phase. In the enrollment phase, the system enrolls a group of $k$ speakers, with speaker identities $\mathcal{U}=\{1,2,\ldots,k\}$. In the test phase, the system aims to recognize the speaker identity of a test voice $\bm{x}$, which has the following two situations:

\textit{CSI:} It classifies $\bm{x}$ into one of the enrolled speakers. The decision module of CSI $D(\bm{x})$ is defined as:
\begin{equation}
    D(\bm{x})=\underset{i \in \mathcal{U}}{\operatorname{argmax}}[S(\bm{x})]_{i}
\end{equation}
where $[S(\bm{x})]_{i}$ denotes the similarity score between $\bm{x}$ and the $i$-th enrollment speaker produced by the speaker identification system.

\textit{OSI:} It either identifies $\bm x$ as one of the enrolled speakers or determines that $\bm x$ does not belong to any of the speakers in $\mathcal{U}$. The decision module of OSI $D(\bm{x})$ is defined as:
\begin{equation}
    D(\bm{x})=\left\{\begin{array}{ll}
    \underset{i \in \mathcal{U}}{\operatorname{argmax}}[S(\bm{x})]_{i}, & \mathrm{ if }\ \underset{i \in \mathcal{U}}{\operatorname{max}}[S(\bm{x})]_{i} \geq \theta  \\
    \mathrm{reject}, & \mathrm{ otherwise}
    \end{array}\right.
\end{equation}
where $\theta$ is a predefined decision threshold $\theta$.

\section{Proposed Method}
\subsection{Problem formulation}
A targeted adversarial attack to the speaker identification system $D(\bm{x})$ aims to generate a perturbation $\bm \delta$ to the input voice $\bm{x}$ of a source speaker $s$, such that the system may misclassify a generated adversarial example $\bm{x}^{\prime} = \bm{x} + \bm \delta$ as a target speaker $t \in \mathcal{U}=\{1, \cdots, k\}$, i.e. $t = D(\bm x^{\prime})$, while the difference between $\bm x$ and $\bm{x}^{\prime}$ is as small as possible and may be unaware by humans. Usually, the perturbation $\bm \delta$ is generated by an adversarial attacker $\bm \delta=G_{\alpha}(\bm x)$, where $\alpha$ is the learnable parameters of the attacker. To design an effective attacker, the core is to design a loss function, denoted as $L$, such that, when $L$ is minimized, $t = D(\bm x^{\prime})$ is achieved.

This paper focuses on adversarial attack techniques in the time domain, where the attacker $G_{\alpha}(\cdot)$ takes the waveform voice $\bm x$ as its input and generates $\bm \delta$ in the time domain. In the following of this section, we describe the proposed attacker, i.e. SSED, in detail.

\subsection{Framework}

The proposed SSED aims to generate adversarial perturbations with high
TASR and high SNR.
The architecture of the proposed SSED is show in Fig. \ref{fig:SSED}. It consists of an encoder $\mathcal{E}$, a perturbation decoder $\mathcal{G}_{\mathcal{N}}$, and a saliency map decoder $\mathcal{G}_{\mathcal{M}}$, where the saliency map decoder is one of the core novelties of the proposed method.


SSED generates the perturbation $\bm \delta$ for the voice of the source speaker $\bm x$ in the following process. First, the encoder $\mathcal{E}$ encodes $\bm x$ into a latent vector $\mathbf{y}$. Then, the perturbation decoder $\mathcal{G}_{\mathcal{N}}$ creates noise $\bm n$ from $\mathbf{y}$, i.e. $\bm n = \mathcal{G}_{\mathcal{N}}(\mathbf{y})$. In the meantime, the saliency map decoder $\mathcal{G}_{\mathcal{M}}$ generates a mask $\bm m$ from $\mathbf{y}$. The final adversarial perturbation $\bm \delta$ is generated by:
\begin{equation}
\bm \delta=\varepsilon (\bm n \odot \bm m)
\end{equation}
where $\varepsilon$ is a manually-tunable scaling factor for controlling the amplitude of the perturbation, and the symbol ``$\odot$'' denotes the dot-product operator. Note that, the mask $\bm m$ is designed to emphasize the important samples of $\bm x$ that affect the decision $D(\bm x)$. Also, $\bm x$, $\bm n$, and $\bm m$ are all voice signals in the time domain. The above modules are all convolutional residual networks. See Section \ref{sec:details} for the detailed settings of the network structures.



SSED jointly trains the above modules by minimizing the following loss function:
\begin{equation}
    L=  L_{\mathrm{SNR}} +  L_{\mathrm{ASR}}\label{con:loss_total}
\end{equation}
where $L_{\mathrm{SNR}}$ and $L_{\mathrm{ASR}}$ are two components for improving
the SNR and TASR of the adversarial examples respectively. In the following two subsections, we describe the two components respectively.

\subsection{Saliency map decoder}
Unlike existing adversarial attack approaches for speaker identification that simply minimizes $L_{\mathrm{norm}}=\left[\max \left(\bm x-\bm x^{\prime}, 0\right)\right]^{2}$ to maximize the SNR \cite{li2020universal}, here we propose an additional saliency map decoder for further improving the SNR. The keyword ``saliency map'', which was originally used in image processing, e.g.\cite{simonyan2013deep}, is used to make SSED attend to highly sensitive regions of $\mathbf{y}$ that affect the decision of $D(\bm x)$. The saliency map decoder minimizes the following objective:
\begin{equation}
    L_{f}=\sqrt{  ({\bm{m}}')^{T}\bm m'}
\end{equation}
where $\bm m'$ is a normalized saliency map vector whose elements are variables in the range of $[0,1]$:
 \begin{equation}
  {\bm m}'=\frac{{\bm m} - \min({\bm m})}{\max({\bm m}) - \min({\bm m})}.
\end{equation}

 Making SSED focus on the highly sensitive regions of the spectral representation $\mathbf{y}$ has the following two merits. First, SSED preserves the spectral characteristics of the original voice $\bm x$, and makes $\bm\delta$ difficult to be detected in the frequency domain. At the same time, the distribution of the perturbation is concentrated to the high energy region of $\bm x$, which makes the adversarial voice $\bm x^{\prime}$ less noisy and more inconspicuous.

 Finally, the SNR of the perturbation is improved by minimizing:
   \begin{equation}
L_{\mathrm{SNR}} = \lambda_f L_{f} + \lambda_{n} L_{\mathrm{norm}}
\label{con:SNR}
\end{equation}
 where $\lambda_f$ and $\lambda_{n}$ are two hyperparameters.

\subsection{Angular loss function}

To improve the TASR of adversarial examples to speaker identification, we need to make the speaker embedding of $\bm x^{\prime}$ as dissimilar as possible to that of the source speaker $s$, and as similar as possible to the target speaker $t$, where a speaker embedding of an utterance is the input of the softmax layer of the speaker identification system. However, existing approaches only explored the latter \cite{xie2021real}, leaving the former far from studied, to our knowledge. In this paper, we propose a new loss function, named angular loss, to enlarge the dissimilarity between the speaker embeddings of $\bm x^{\prime}$ and the source speaker $s$.

Specifically, we denote the speaker embedding of $\bm x^{\prime}$ as $\mathbf{z}^{\prime}$, and the speaker embedding of $\bm x$ as $\mathbf{z}$. The angular loss minimizes:
\begin{equation}
    L_{\mathrm{angular}}=\frac{\mathbf{z}^T \mathbf{z}^{\prime}}{\max \left(\|\mathbf{z}\|_2 \left\|\mathbf{z}^{\prime}\right\|_2, \epsilon\right)}.
    \label{con:arg}
\end{equation}
where $\epsilon$ is a small constant, which is set to $10^{- 12}$ in this paper.

We also adopt the conventional speaker loss \cite{chen2021real,zhang2022imperceptible} as part of $L_{\mathrm{ASR}}$. It has different forms for CSI and OSI respectively. For the integrity of the paper, we present them briefly as follows. The speaker loss for CSI is defined as:
\begin{equation}
    L_{\mathrm{speaker}} = \max _{i \in \mathcal{U}, i \neq t}[S(\bm x^{\prime})]_{i}-[S(\bm x^{\prime})]_{t}
    \label{con:CSI_t_s}
\end{equation}
which aims to make the targeted speaker identification system wrongly identify $\bm x^{\prime}$ as the target speaker $t$. The speaker loss for OSI also have to guarantee that the decision score $[S(\bm x^{\prime})]_{t}$ is larger than the threshold $\theta$:
\begin{equation}
     L_{\mathrm{speaker}} = \max \left\{\max _{i \in \mathcal{U}, i \neq t}[S(\bm x^{\prime})]_{i}, \theta\right\}-[S(\bm x^{\prime})]_{t}
    \label{con:OSI_t_s}
\end{equation}

 Finally, the TASR of SSED can be improved by minimizing:
   \begin{equation}
L_{\mathrm{ASR}} = \lambda_{s} L_{\text{speaker}} + \lambda_{a} L_{\mathrm{angular}}
\label{con:ASR}
\end{equation}
 {where $\lambda_{s}$ and $\lambda_{a}$ are two hyperparameters.} Substituting \eqref{con:SNR} and \eqref{con:ASR} into \eqref{con:loss_total} derives the objective of the proposed SSED.

\section{Experiments}
\subsection{Experimental setup}
\subsubsection{Dataset}
We used VoxCeleb1 and VoxCeleb2 datasets \cite{nagrani2020voxceleb}, where the development set of VoxCeleb2 was used for the training of the speaker identification system. All comparison methods were evaluated in both the CSI and OSI scenarios. For CSI, the system was enrolled by 1211 speakers from the development set of Voxceleb1. In the evaluation process, we randomly chose one of the enrolled speakers as the targeted speaker of adversarial attacks. We selected 35 utterances from each speaker of the development set of Voxceleb1 to train adversarial attackers, and another 5 utterances of each speaker as imposters for testing.

 For OSI, the system is enrolled by 5 random speakers from the development set of VoxCeleb2 \cite{chen2021real}. In the evaluation process, we randomly chose one of the 5 enrolled speakers as the targeted speaker. We trained each adversarial attacker with 10 utterances per speaker from the 5994 speakers of the development set of VoxCeleb2, and
 chose 40 speakers from the test set of VoxCeleb1 as imposters to attack the OSI system.


\subsubsection{Targeted speaker identification system}
The Fast ResNet-34 \cite{chung2020defence,9739948} trained with the AAM-Softmax \cite{deng2019arcface} objective was used as the targeted speaker identification system. {We used voxceleb\_trainer\footnote{\url{https://github.com/clovaai/voxceleb_trainer}} to train the system.}
To verify that the targeted system is a state-of-the-art system, we first evaluated the system on the Voxceleb1 development set which consists of 1211 speakers. The system achieves a classification accuracy of 93.15\%. We further transformed the speaker identification system to a verification system by using the cosine scoring back-end, and evaluated the verification system on the VoxCeleb1 Original trial list\footnote{\url{https://www.robots.ox.ac.uk/~vgg/data/voxceleb/meta/veri_test.txt}} that consists of 37k trials from 40 speakers. The EER of the system is 2.34\%.

\subsubsection{Comparison methods}
\label{sec:details}

\renewcommand\arraystretch{1.05}
\begin{table}[t]
  \centering
  \caption{Comparison results on the CSI task. The performance of the top two methods are marked in bold.}
  \small
  \scalebox{0.8}{
    \begin{tabular}{c|c|c|c|c}
    \hline
    Attack Type & $\varepsilon$ & TASR (\%) & SNR (dB)   & Time (s) \\
    \hline
    None  &  -    & 0.09  &  -    &  - \\
    \hline
    \multirow{4}[2]{*}{FGSM} & 0.001 & 5.42  & 41.40  & \multirow{4}[2]{*}{0.9} \\
         & 0.002 & 10.03 & 35.42 &  \\
         & 0.005 & 14.91 & 27.47 &  \\
         & 0.01  & 16.09 & 21.45 &  \\
    \hline
    \multirow{3}[2]{*}{BIM-10} & 0.001 & 63.68 & 43.49 & \multirow{3}[2]{*}{\textbf{5.86}} \\
         & 0.002 & 90.23 & 38.48 &  \\
         & 0.005 & \textbf{99.39} & \textbf{31.65} &  \\
    \hline
    C\&W-L2  &  -    & 52.88 & 63.97 & 130.21 \\
    \hline
    UAPs \cite{li2020universal}  &  - & 65.6  & 38.56  & 0.004 \\
    \hline
    \multirow{3}[2]{*}{SSED (proposed)} & 0.01  & 45.4  & 52.07 & \multirow{3}[2]{*}{\textbf{0.41}} \\
         & 0.03  & 90.0  & 43.19 &  \\
         & 0.05  & \textbf{97.3}  & \textbf{39.07} &  \\
    \hline
    \end{tabular}%
    }
  \label{tab:CSI}%
\end{table}%

\begin{table}[t]
  \centering
  \caption{Comparison results on the OSI task.}
  \small
    \scalebox{0.8}{
    \begin{tabular}{c|c|c|c|c}
    \hline
    Attack Type & $\varepsilon$ & TASR (\%) & SNR (dB)   & Time (s) \\
    \hline
    None  &  -    & 0.84  &  -    &  - \\
    \hline
    \multirow{4}[2]{*}{FGSM} & 0.001 & 15.61 & 41.42 & \multirow{4}[2]{*}{0.14} \\
          & 0.002 & 21.5  & 35.40  &  \\
          & 0.005 & 21.34 & 27.44 &  \\
          & 0.01  & 16.45 & 21.42 &  \\
    \hline
    \multirow{3}[2]{*}{BIM-10} & 0.001 & 77.9  & 43.42 & \multirow{3}[2]{*}{\textbf{1.16}} \\
          & 0.002 & 95.9  & 38.42 &  \\
          & 0.005 & \textbf{99.79} & \textbf{31.68} &  \\
    \hline
    C\&W-L2  &  -    & 93.66 & 64.02 & 9.25 \\
    \hline
    UAPs \cite{li2020universal}  &  - & 56.0  & 33.66  & 0.004 \\
    \hline
    \multirow{3}[2]{*}{SSED (proposed)} & 0.01  & 74.8  & 51.88 & \multirow{3}[2]{*}{\textbf{0.41}} \\
         & 0.03  & 92.8  & 43.1 &  \\
         & 0.05  & \textbf{98.0}  & \textbf{39.2} &  \\
    \hline
    \end{tabular}%
    }
  \label{tab:OSI}%
\end{table}%

\renewcommand\arraystretch{1.05}
\begin{table}[t]
  \centering
  \caption{Performance comparison of the SSED with or without the saliency map decoder. }
  \small
  \scalebox{0.8}{
    \begin{tabular}{c|c|c|c}
    \hline
    Task  & Type  & TASR (\%) & SNR (dB) \\
    \hline
    \multirow{2}[2]{*}{CSI} & with saliency map decoder & \textbf{97.3} & \textbf{39.07} \\
          & without saliency map decoder & 95.1  & 35.62 \\
    \hline
    \multirow{2}[2]{*}{OSI} & with saliency map decoder & \textbf{98.0} & \textbf{39.2} \\
          & without saliency map decoder & 96.7    & 37.1 \\
    \hline
    \end{tabular}
    }%
  \label{tab:w/wo saliency}%
\end{table}%

\begin{figure}[t]
\centering
\subfigure[Waveform]{
	\centering
	\label{fig:wav}
	\includegraphics[width=0.21\textwidth]{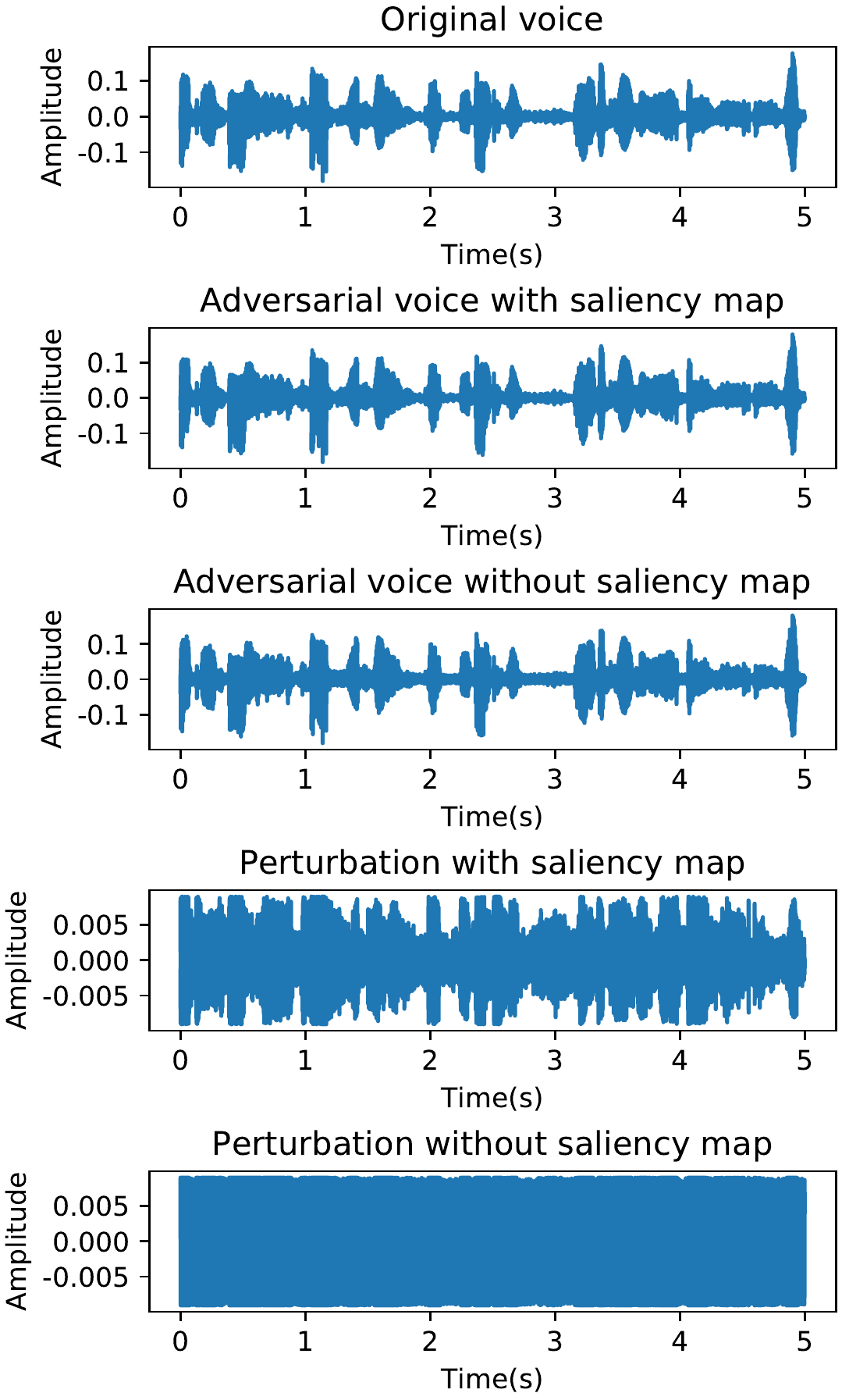}
}\subfigure[Spectrogram]{
	\centering
	\label{fig:spectrogram}
	\includegraphics[width=0.21\textwidth]{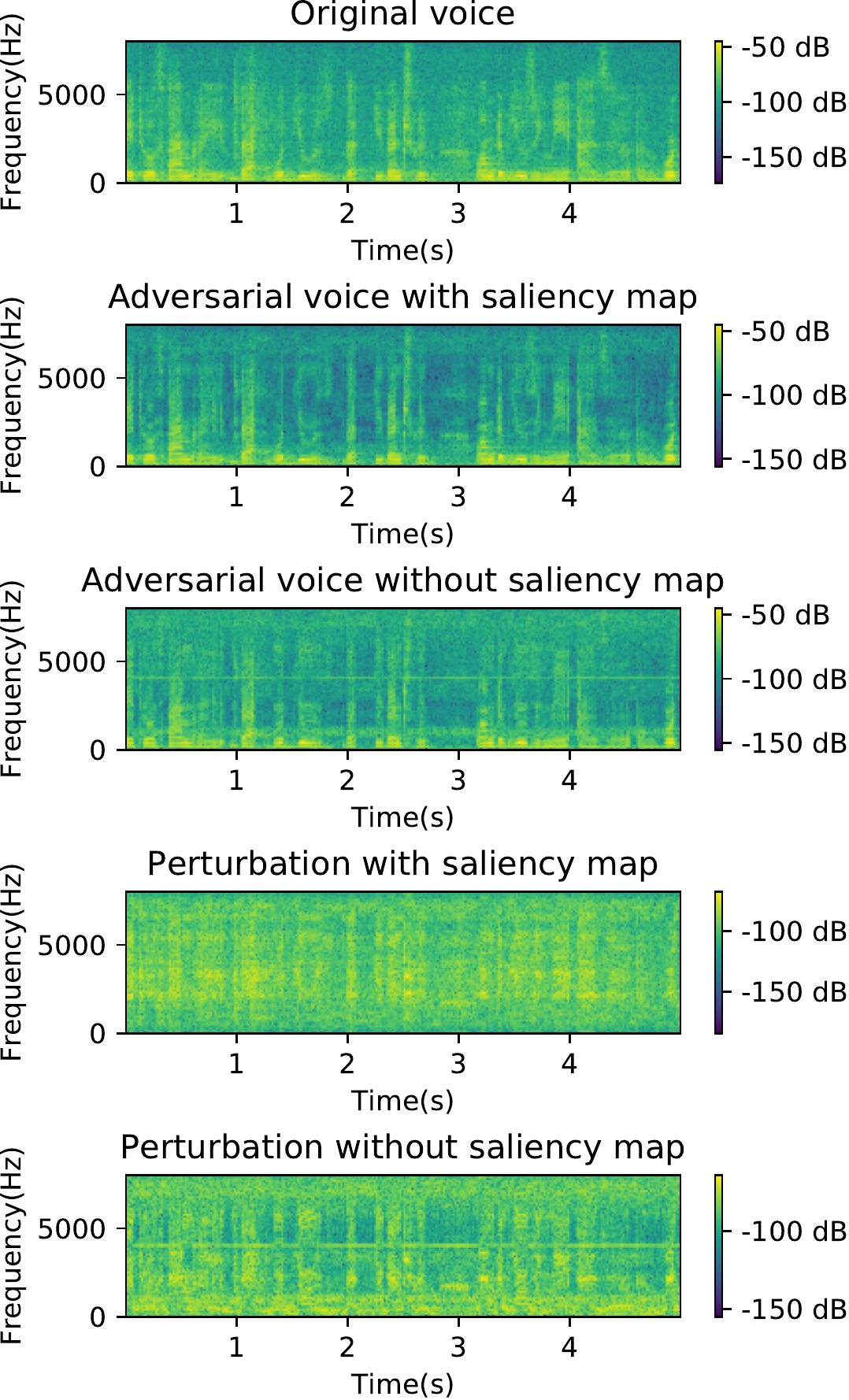}
}
\caption{
    A comparison example between the original voice and its adversarial examples generated by either SSED or the SSED variant without the saliency map decoder.
    The SNR of the adversarial examples generated with and without the saliency map decoder are 51.79dB and 43.47dB, respectively.}
	\label{fig:spec}
\end{figure}

\renewcommand\arraystretch{1.05}
\begin{table}[t]
  \centering
  \caption{Performance comparison of the SSED with or without the angular loss.}
  \small
  \scalebox{0.8}{
    \begin{tabular}{c|c|c|c}
    \hline
    Task  & Type  & TASR (\%) & SNR (dB) \\
    \hline
    \multirow{2}[2]{*}{CSI} & with angular loss & \textbf{97.3} & \textbf{39.07} \\
          & without angular loss & 96.8  & 38.85 \\
    \hline
    \multirow{2}[2]{*}{OSI} & with angular loss & \textbf{98.0} & \textbf{39.2} \\
          & without angular loss & 96.9    & 38.92 \\
    \hline
    \end{tabular}%
    }
  \label{tab:w/wo angular}%
\end{table}%

\begin{figure}[t]
\begin{center}
\includegraphics[width=0.41\textwidth]{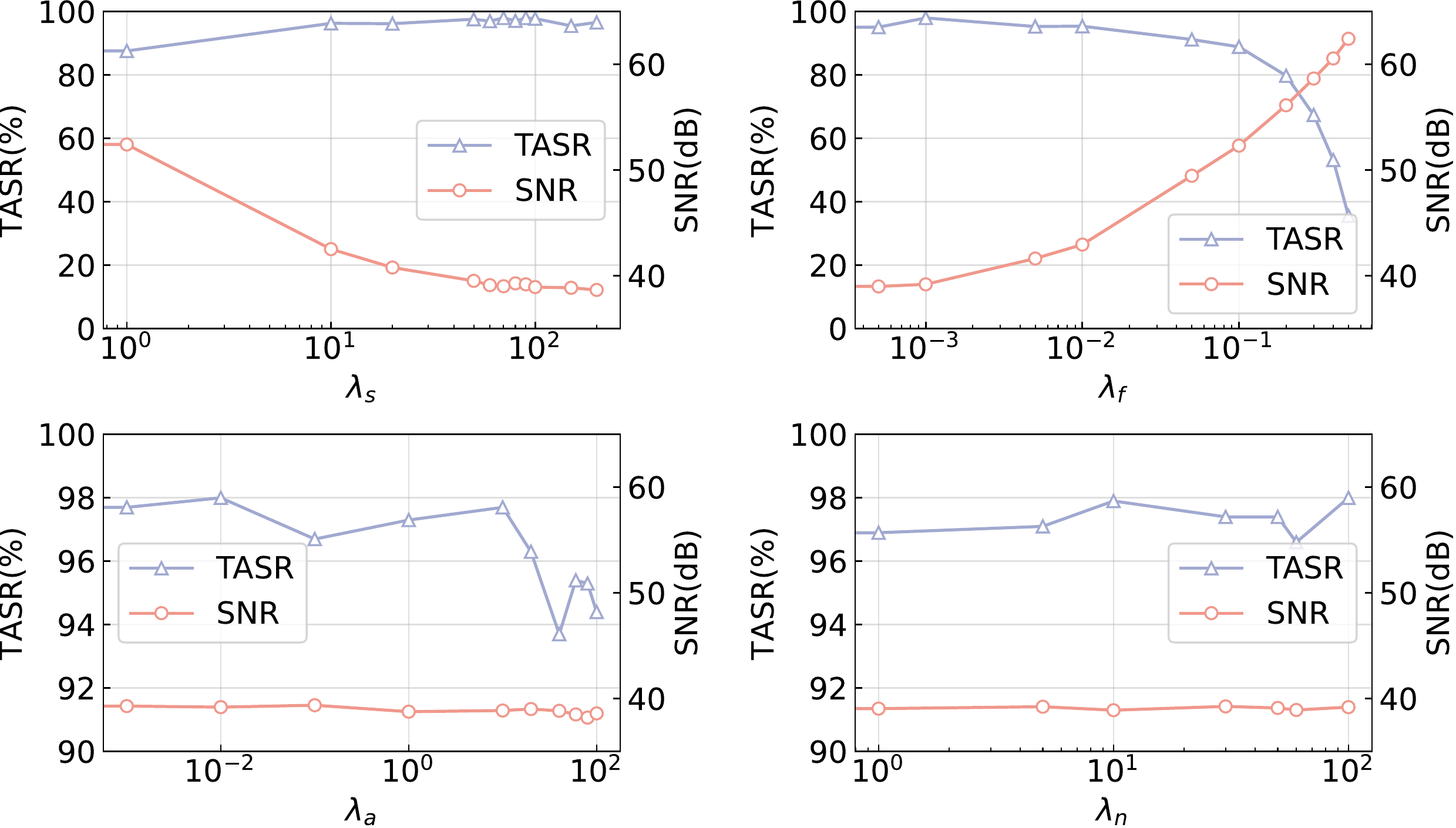}
\end{center}
\caption{Effects of the hyper-parameters on the OSI task.}
\label{fig:hp_OSI}
\end{figure}

The parameter settings of the proposed SSED are as follows. The encoder $\mathcal{E}$ consists of 3 convolution blocks and 6 residual blocks. The 1-D convolution, batch normalization, and ReLU were applied in every convolution block. The kernel sizes for all convolution layers were set to 7, 3, and 3, respectively. The perturbation decoder $\mathcal{G}_{\mathcal{N}}$ consists of 3 transposed convolution blocks, whose kernel sizes are 3, 3, and 7, respectively. The saliency map decoder $\mathcal{G}_{\mathcal{M}}$ has a similar structure with
$\mathcal{G}_{\mathcal{N}}$ except the last transposed convolution layer. We trained SSED by 10 epochs with the Adam optimizer and a learning rate of $10^{-3}$.

We compared the proposed method with four well-known adversarial attack methods which are described as follows. FGSM \cite{goodfellow2014explaining} is a gradient-based one-step $l_{\infty}$ attack. BIM-10 \cite{kurakin2016adversarial} is an iterative version of FGSM with 10 iterations. C$\&$W \cite{carlini2017towards} is an optimization-based attack. UAPs \cite{li2020universal} is a universal generation-network-based attack approach.

\subsubsection{Evaluation metrics}
The first evaluation metric is TASR. It refers to the probability of identifying an imposter voice as the target speaker by the targeted speaker identification system. The second evaluation metric is SNR, which is defined as $\operatorname{SNR}=10 \log _{10}\left(P_{x} / P_{\delta}\right)$ where $P_{x}$ and $P_{\delta}$ are the signal power of the original voice $\bm x$ and the power of the perturbation $\bm \delta$ respectively. Finally, in order to evaluate the efficiency, we also recorded the time for generating the adversarial examples.

\subsection{Results}
\subsubsection{Main results}
Tables \ref{tab:CSI} and \ref{tab:OSI} list the comparison results of the proposed SSED with the four referenced methods on the CSI and OSI tasks respectively. From the tables, we see that the proposed SSED is comparable to BIM-10, and significantly outperforms the other comparison methods, if we take an overall consideration of TASR and SNR. However, SSED needs much less time to generate the perturbations. Comparing to UAPs which is also a generation-network-based approach, SSED yields higher TASR and SNR than UAPs, with the expense of higher computational complexity. The phenomenon demonstrates the effectiveness of the novel points of SSED.

\subsubsection{Effect of the saliency map decoder on performance}
We compared SSED with a variant of SSED that removes the saliency map decoder. From the comparison results in Table \ref{tab:w/wo saliency}, we see that the saliency map decoder leads to higher TASR and SNR. Fig. \ref{fig:spec} gives an example on how the saliency map decoder affects adversarial examples. From the figure, we see that both the waveform and the spectrogram of the adversarial example produced by SSED are more similar to the original voice than those produced by the SSED variant, while the perturbation produced by SSED shows lower global energy and clearer local patterns than that produced by the SSED variant. These phenomena are caused by that the saliency map decoder assigns different weights to different samples according to their importance to speaker identification.

\subsubsection{Effect of the angular loss on performance}
Table \ref{tab:w/wo angular} lists the results of SSED and its variant without the angular loss. From the table, we see that the angular loss leads to better performance in terms of both TASR and SNR.

\subsubsection{Effects of hyperparameters on performance}

We tuned the hyperparameters $\lambda_{s}$, $\lambda_{f}$, $\lambda_{a}$ and $\lambda_{n}$ to achieve a trade-off between TASR and SNR. To evaluate their effects on performance, we tuned each hyperprameter respectively on the OSI task. For tuning each hyperparameter, we fixed the other three to their default values. From the results in Fig. \ref{fig:hp_OSI}, we see that $\lambda_{f}$, which controls the importance of the saliency map decoder, is of great help to the improvement of SNR, with a negative effect on TASR. $\lambda_{s}$, which controls the speaker loss, is important to the improvement of TASR, with a negative effect on SNR. $\lambda_{a}$ and $\lambda_{n}$ have slight effects on TASR and SNR.
Note that, similar phenomena were observed on the CSI task as well.

\section{Conclusions}
In this paper, we propose the SSED model to generate adversarial attack against speaker identification. SSED adopts a novel saliency map decoder to assign different weights to the samples of an utterance according to their importance to the decision of the targeted speaker identification system, and further used a novel angular loss to push the adversarial example of the utterance away from its source speaker. Our experimental results on both the CSI and OSI tasks demonstrate the effectiveness of the two novel points, which lead to the state-of-the-art performance of SSED with a low computational cost for generating the adversarial examples.

\bibliographystyle{IEEEtran}
\bibliography{ref}
\end{document}